\documentclass[final,english]{bullsrsl}[2022/06/15]

\usepackage[latin1]{inputenc}
\usepackage[T1]{fontenc}
\usepackage{natbib} 
\usepackage{graphicx}
\usepackage{xcolor}

\begin{document}

\title{Survey of Variables with the ILMT}

\author[affil={1}, corresponding]{Baldeep}{Grewal}
\author[affil={2,3}]{Bhavya}{Ailawadhi}
\author[affil={4,5}]{Talat}{Akhunov}
\author[affil={6}]{Ermanno}{Borra}
\author[affil={2,7}]{Monalisa}{Dubey}
\author[affil={2,7}]{Naveen}{Dukiya}
\author[affil={1}]{Jiuyang}{Fu}
\author[affil={1}]{Paul}{Hickson}
\author[affil={2}]{Kuntal}{Misra}
\author[affil={2}]{Brajesh}{Kumar}
\author[affil={2,3}]{Vibhore}{Negi}
\author[affil={2,8}]{Kumar}{Pranshu}
\author[affil={1}]{Ethen}{Sun}
\author[affil={9}]{Jean}{Surdej}
\affiliation[1]{Department of Physics and Astronomy, University of British Columbia, 6224 Agricultural Road, Vancouver, BC V6T 1Z1, Canada}
\affiliation[2]{Aryabhatta Research Institute of Observational sciencES (ARIES), Manora Peak, Nainital, 263001, India}
\affiliation[3]{Department of Physics, Deen Dayal Upadhyaya Gorakhpur University, Gorakhpur, 273009, India}
\affiliation[4]{National University of Uzbekistan, Department of Astrophysics, 100174 Tashkent, Uzbekistan}
\affiliation[5]{ Ulugh Beg Astronomical Institute of the Uzbek Academy of Sciences, Astronomicheskaya 33, 100052 Tashkent, Uzbekistan}
\affiliation[6]{Department of Physics, Universit\'{e} Laval, 2325, rue de l'Universit\'{e}, Qu\'{e}bec, G1V 0A6, Canada}
\affiliation[7]{Department of Applied Physics, Mahatma Jyotiba Phule Rohilkhand University, Bareilly, 243006, India}
\affiliation[8]{Department of Applied Optics and Photonics, University of Calcutta, Kolkata, 700106, India}
\affiliation[9]{Institute of Astrophysics and Geophysics, University of Li\`{e}ge, All\'{e}e du 6 Ao$\hat{\rm u}$t 19c, 4000 Li\`{e}ge, Belgium}

\correspondance{grewal.16@hotmail.com}
\date{May 15, 2023}
\maketitle


%

\begin{abstract}
Nestled in the mountains of Northern India, is a 4-metre rotating dish of liquid mercury. Over a 10-year period, the International Liquid Mirror Telescope (ILMT) will survey 117 square degrees of sky, to study  the astrometric and photometric variability of all detected objects. One of the scientific programs will be a survey of variable stars. The data gathered will be used to construct a comprehensive catalog of light curves. This will be an essential resource for astronomers studying the formation and evolution of stars, the structure and dynamics of our Milky Way galaxy, and the properties of the Universe as a whole. This catalog will be an aid in our advance to understanding the cosmos and provide deeper insights into the fundamental processes that shape our Universe. In this work, we describe the survey and give some examples of variable stars found in the early commissioning data from the ILMT.
\end{abstract}


\section{Introduction}
Variable stars are of fundamental importance to many areas of astronomy, including distance measurements, stellar structure and evolution, studies of compact objects, and close binary systems. Variables are any class of object whose brightness changes with time, some exhibit periodicity while others do not. Periodic variables are especially important because the prediction of their magnitudes can be known in the future. Variable objects, such as stars, eclipsing binaries, pulsators, and novae are scattered throughout the night sky. They provide valuable information on the evolution and nature of stars in our Milky Way. Cepheids and RR Lyrae serve as standard candles, where distances can be calculated once their absolute magnitude is known. Cepheid Variables provide crucial information in determining distances and the age of the Universe.  Studies of accretion disks in cataclysmic variables (CVs) can help understand activity inside active galaxies with supermassive black holes. Studies of variable stars  \citep[e.g. ][]{Pietrukowicz_2009} have the potential to illuminate many areas of astrophysics. 

In northern India, Devasthal Peak ($79^{\circ} 41' 04''$ E, $29^{\circ} 21' 40''$ N) at an altitude of 2450 m, hosts several optical facilities including the 4-m, zenith-pointing, International Liquid Mirror Telescope \citep[ILMT,][]{Surdej_2018}. The ILMT is dedicated to performing a broad-band photometric survey with a focus on variability studies. This unique telescope employs a rotating mirror that uses liquid mercury as its reflecting surface. The mirror rotates with a period of 8.02 sec, which is controlled to an accuracy of a few parts per million. It images the region of sky passing overhead onto a $4096 \times 4096$-pixel charge-coupled device (CCD) that operates in time-delay integration mode \citep{Gibson_1992}. This allows it to obtain near-continuous imaging of the sky passing overhead, with an integration time of 102.4 s (the time taken for an astronomical object to cross the CCD due to the rotation of the Earth). A five-element optical corrector removes both telescope distortion and star-trail curvature. The telescope has a $22.4' \times  22.4'$ field of view and is equipped with Sloan Digital Sky Survey (SDSS) g$'$, r$'$, and i$'$ filters \citep{Kumar_2022}. The main scientific objectives of the ILMT are studies of astrometric and photometric variability of all detectable objects in the 117 square degrees of sky accessible to the telescope. 

\section{Observations and Analysis}

\begin{figure}[t]
\centering
\includegraphics[width=0.7\textwidth]{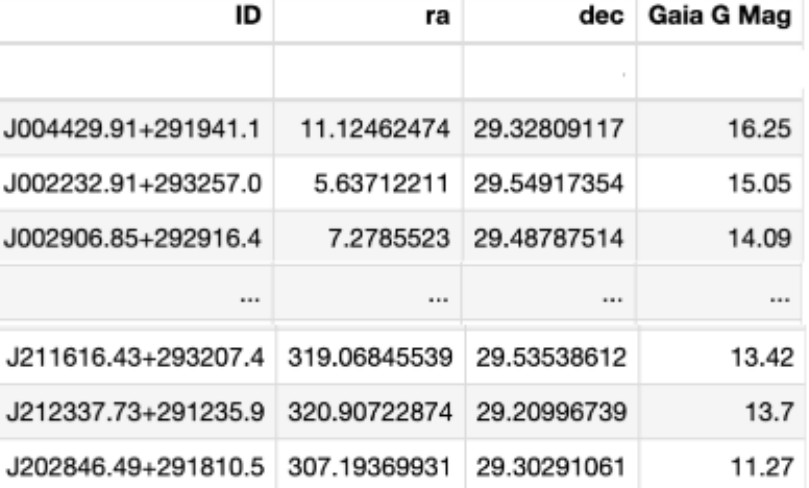}
%
\begin{minipage}{12cm}
\caption{A concise table showcasing the structure of the CSV file. This is for all the stars that are in the ILMTs imaging area. }
\label{fig:csv}
\end{minipage}
\end{figure}

The All-Sky Automated Survey for Supernovae (ASAS-SN) catalog of variables stars \citep{Christy_2022} is a useful starting point for variable star studies. From the data available on the ASAS-NS website, a sub-catalog was created that contains data for all objects within the ILMT survey area. The main selection criteria for the objects selected was that they are in the field of view of the ILMT. Variable stars with long and short periods are considered. With the long-term observations of the ILMT, enough data can be collected spanning multiple periods. Objects were grouped according to the types of variable: Cepheids, RR Lyrae, Mira, Rotational, Eclipsing Binaries, and Semi-Regular/Irregular. The sub-catalog data are in the form of comma-separate variable (CSV) files, with fields (i) ASAS-SN name (ii) right ascension (J2000) (iii) declination (J2000), and (iv) Gaia G magnitude. These files were then converted to FITS format using Astropy. An example of the CSV file is showcased in Fig.\,\ref{fig:csv}. 

\begin{figure}[t]
\centering
\includegraphics[width=0.7\textwidth]{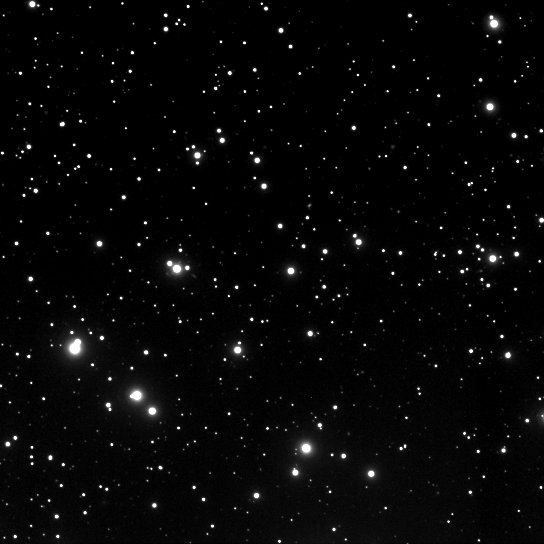}
\bigskip

\begin{minipage}{12cm}
\caption{An example image from one night of ILMT observations. This image was taken on the night of October 31, 2022.}
\label{fig:ILMT}
\end{minipage}
\end{figure}

The observational data consisted of images obtained on November 1 and 2 2022, along with images in early March 2023, during its commissioning phase. Fig.\,\ref{fig:ILMT} showcases the ILMTs imaging abilities which were taken during this early commissioning phase. Although this is a small sample, it is sufficient to allow the analysis techniques to be developed and the performance investigated. These images were preprocessed and an initial astrometric and photometric calibration was performed based on Gaia stars in the field. The typical astrometric accuracy is 0.3 arcsec and the photometric accuracy ranges from 0.05 to 0.10 magnitude, on clear nights.

A Python photometry program \texttt{SunPhot} \citep{SunEProceedings} was then used to extract the magnitude of the cataloged objects present in the ILMT images. The right ascension and declination of the stars in the catalog were matched to the calibrated ILMT science images. Once an object has been found, a circular aperture, centred on the object, and a concentric annulus was defined. The purpose of the annulus is to calculate the local background around the star. The median intensity within the annulus provides an estimate of the sky background, which was then subtracted from each pixel in the aperture to compute the object's flux. The fluxes were then converted to magnitudes using the photometric zero point derived from the Gaia stars in the image. 

\section{Results}
\begin{figure}[t]
\centering
\includegraphics[width=\textwidth]{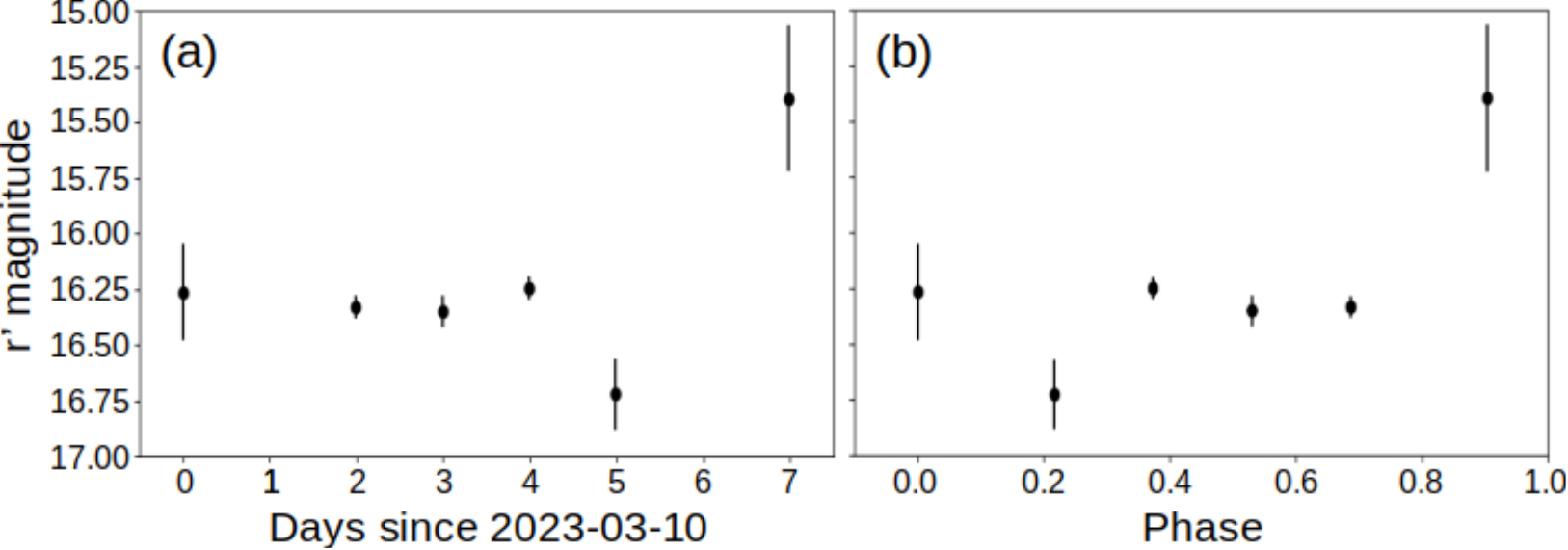}
\bigskip

\begin{minipage}{12cm}
\caption{RR Lyrae variable ASASSN-V J140021.26+292506.1 (UU CVn), observed in the March 2023 ILMT data. Reference Epoch 2460018 JD.(a) Observed light curve. This star has a pulsation period of 0.541 days \citep{DR2,DR2VisieR}. When sampled by the ILMT with a period of 1 sidereal day, the accumulated phase difference is 0.94 over the 7-day interval shown. (b) Light curve obtained by folding the data at the 0.541 period. }
\label{fig:RRLyrae}
\end{minipage}
\end{figure}

There are 417 known variable stars in the ILMT field of view. In our limited data set, 23 RR Lyrae variables were observed, out of a total number of 69 in the full ILMT survey area. These variables typically have periods of 0.2 to 2 days and will be undersampled by the 1-day cadence of the ILMT. Nevertheless, for periodic variables, a light curve can be built up from many nights of observations, by folding the data at the period of the variable. This procedure can also be used to determine the period, by minimizing the spread of values in the folded data. As an example, Fig.\,\ref{fig:RRLyrae} shows measurements of an RR Lyrae star that has a period of 0.541 days. Although, more data are needed to produce a robust phase plot and reliably estimate the photometric uncertainties. Nevertheless, Fig.\,3 showcases that the ILMT can provide a phase plot even for variables whose period is shorter than one day. One, therefore, expects little change in magnitude when the star is observed by the ILMT on the next night, 0.997 days (one sidereal day) later, as the star is seen at nearly the same phase. However daily observations over a week sample the full range of phase, and additional observations will improve the sampling of the folded light curve (Fig.\,\ref{fig:RRLyrae}b).

A total of 73 rotational variables are in the ILMT survey area and 31 have been observed so far. These rotational variables present an excellent opportunity to investigate the surface characteristics of stars by examining the temporal variations in their brightness. Furthermore, there are 11 Mira variables and only one has been observed to date. 

The last type of variable in this study is Semi-Regular/Irregular stars. These show variation changes on timescales of months to years that do not appear to be periodic.  Currently, 40 have been observed by the ILMT, out of a total of 73 in the full survey area. 

\section{Discussion}

Cepheid variables have long been used for distance measurements by virtue of their period-luminosity (PL) relationship. ILMT observations can in principle provide high-accuracy period measurements. Combined with Gaia parallaxes, this will allow uncertainties in the PL relation to be reduced.  RR Lyrae stars follow a period-colour-luminosity relationship and can thus serve as independent distance indicators. They are less bright than Cepheids, but they can still be useful indicators for determining the distances to nearby galaxies.

The ILMT has the potential to detect Cepheids and RR Lyrae variables at significant distances, but the actual detection limits depend on multiple factors, such as the intrinsic brightness of the star and the observing conditions.

 As the ILMT is still in the commissioning stage, the amount of data available at this time is limited but growing steadily.  Improvements to the telescope alignment and mirror balance have resulted in image quality that is now approaching 1.4 arcsecs (full width at half maximum intensity).  Further improvement is expected as the team gains experience with the telescope and its systems. We have demonstrated the performance of the ILMT, and expect that the survey that it is conducting will be a useful resource for studies of variable stars, galaxies, quasars, and gravitational lenses. 

\begin{acknowledgments}
The 4m International Liquid Mirror Telescope (ILMT) project results from a collaboration between the Institute of Astrophysics and Geophysics (University of Li\'{e}ge, Belgium), the Universities of British Columbia, Laval, Montreal, Toronto, Victoria and York University, and the Aryabhatta Research Institute of observational sciencES (ARIES, India). The authors thank Hitesh Kumar, Himanshu Rawat, Khushal Singh and other observing staff for their assistance at the 4m ILMT.  The team acknowledges the contributions of ARIES's past and present scientific, engineering and administrative members in the realisation of the ILMT project. JS wishes to thank Service Public Wallonie, F.R.S.--FNRS (Belgium) and the University of Li\'{e}ge, Belgium for funding the construction of the ILMT. PH acknowledges financial support from the Natural Sciences and Engineering Research Council of Canada, RGPIN-2019-04369. PH and JS thank ARIES for hospitality during their visits to Devasthal. BA acknowledges the Council of Scientific $\&$ Industrial Research (CSIR) fellowship award (09/948(0005)/2020-EMR-I) for this work. MD acknowledges Innovation in Science Pursuit for Inspired Research (INSPIRE) fellowship award (DST/INSPIRE Fellowship/2020/IF200251) for this work. This work is supported by the Belgo-Indian Network for Astronomy and astrophysics (BINA), approved by the International Division, Department of Science and Technology (DST, Govt. of India; DST/INT/BELG/P-09/2017) and the Belgian Federal Science Policy Office (BELSPO, Govt. of Belgium; BL/33/IN12)

\end{acknowledgments}

\begin{furtherinformation}

\begin{authorcontributions}
This work results from a long-term collaboration in which all authors have made significant contributions. 
\end{authorcontributions}

\begin{conflictsofinterest}
The authors declare no conflict of interest. 
\end{conflictsofinterest}

\end{furtherinformation}

\bibliographystyle{bullsrsl-en}
\bibliography{S11-P16_GrewalB}

\end{document}